\documentclass{llncs}
\usepackage[T1]{fontenc}
\usepackage{graphicx}

\usepackage{amsmath}
\usepackage{amssymb}
\usepackage{xpatch}

\makeatletter
\newinsert\copyrightnote@ins
\count\copyrightnote@ins=1000
\dimen\copyrightnote@ins=8in
\newcommand*\copyrightnote@hook
  {%
    \unvbox\copyrightnote@ins
    \global\let\@makecol\copyrightnote@makecol
  }
\let\copyrightnote@AtBeginDocument\AtBeginDocument
\AtBeginDocument{\let\copyrightnote@AtBeginDocument\@firstofone}
\newcommand*\copyrightnote@firstuse
  {%
    \gdef\copyrightnote@firstuse
      {\global\skip\copyrightnote@ins=\bigskipamount\gdef\copyrightnote@firstuse{}}%
    \global\let\copyrightnote@makecol\@makecol
    \xpatchcmd\@makecol{\unvbox\footins}{\unvbox\footins\copyrightnote@hook}
      {}{\GenericError{}{patching @makecol failed}{}{}}
    \copyrightnote@AtBeginDocument
      {%
        \ifvoid\footins
          \insert\footins{}%
        \else
          \global\skip\copyrightnote@ins=\bigskipamount
        \fi
      }%
  }
\newcommand\copyrightnote[1]
  {%
    \copyrightnote@firstuse
    \copyrightnote@AtBeginDocument
      {%
        \insert\copyrightnote@ins
          {%
            \reset@font\footnotesize
            \interlinepenalty\interfootnotelinepenalty
            \splittopskip\footnotesep
            \splitmaxdepth\dp\strutbox
            \floatingpenalty\@MM
            \hsize\columnwidth
            \@parboxrestore
            \color@begingroup
            \parindent1em
            \noindent
            \hskip1.8em
            \ignorespaces #1\@finalstrut\strutbox
            \color@endgroup
          }%
      }%
  }
\makeatother

\copyrightnote{\copyright2022 SPRINGER NATURE. Personal use of this material is permitted. Permission from IEEE must be obtained for all other uses, in any current or future media, including reprinting/republishing this material for advertising or promotional purposes, creating new collective works, for resale or redistribution to servers or lists, or reuse of any copyrighted component of this work in other works..\hspace{\columnsep}\makebox[\columnwidth]{ }}

\begin{document}
\title{How to Extend the Abstraction Refinement Model for Systems with Emergent Behavior?}
%
%
\author{Mohamed Toufik Ailane \and  Christoph Knieke \and Andreas Rausch}
\authorrunning{A. Mohamed Toufik et al.}
%
\institute{Institute for Software and Systems Engineering, Clausthal University of Technology, Clausthal-Zellerfeld, Germany
\email{\{mohamed.toufik.ailane,christoph.knieke,andreas.rausch\}@tu-clausthal.de}\\
\url{https://www.isse.tu-clausthal.de}}
\maketitle             

\begin{abstract}
The Abstraction Refinement Model has been widely adopted since it was firstly proposed many decades ago. This powerful model of software evolution process brings important properties into the system under development, properties such as the guarantee that no extra behavior (specifically harmful behaviors) will be observed once the system is deployed. However, perfect systems with such a guarantee are not a common thing to find in real world cases, anomalies and unspecified behaviors will always find a way to manifest in our systems, behaviors that are addressed in this paper with the name ``emergent behavior''. In this paper, we extend the Abstract Refinement Model to include the concept of the emergent behavior. Eventually, this should enable system developers to: (i) Concretely define what an emergent behavior is, (ii) help reason about the potential sources of the emergent behavior along the development process, which in return will help in controlling the emergent behavior at early steps of the development process.


\end{abstract}

\section{Introduction}


A proliferation of research works have been dedicated to define, formalize and control the ``emergent behavior'' as it can be detrimental - in a way that it can endanger human lives in the case of safety-critical systems, as well as it can be beneficial - in a way that it can fulfill some users requirements or bring some important properties to the system in hand. In this context, one approach to tackle the emergent behavior issues is to consider the behavior during the early stages of the development  of the system as well as during the operational phases.

Different approaches, paradigms and life-cycles were contributed to the wealth of knowledge of software systems development since the early days of software engineering. 
Formal development methods stand as one of the most reliable approaches for software development although the mathematical complexity it might impose. Nevertheless, formal methods gained such excellent worldwide reputation thanks to the ease of correctness and consistency checking. A \textit{correct by construction} system is a fundamental property when it comes to resilient safety-critical systems. 

For the sake of investigating the emergent behavior in software systems, we adopt a model-based development process which is the \textit{waterfall development model} \cite{royce1970software} as a basis of our work to define and formalize the emergent behavior. The waterfall development model is one of the earliest and simplest models in the history of software development, and it is chosen here as an attempt to eliminate any kind of complexity that comes along with the development model, and keep the focus on the concept of the emergence phenomenon and it's relative complexity. As the developer(s) move from one phase to the next one, different models are generated based on each phase. Hence, in order to track the evolution of the software system models, the evolution model \textit{Abstraction Refinement Model} (ARM) is chosen. By tracking the evolution of the system from the very beginning, we investigate how would an emergent behavior emerge or manifest and how does it evolve and finally be observed once the system is deployed.

For this reason, we structure the remainder of this paper as follows: Section~\ref{Section:RelatedWork} will provide some related work regarding the study of the emergence phenomenon in software engineering. Next, Section~\ref{Section:Concepts and definitions} brings some fundamentals, in particular the definition of the \textit{Abstraction Refinement Model} (ARM). 
Our approach towards extending the ARM for systems with emergent behavior is described in Section~\ref{Section:Extended ARM}. Finally, Section~\ref{sec:conclusion} concludes.  

\section{Related Work} \label{Section:RelatedWork}

Emergent behavior in software systems engineering has been a subject of study for a long time. Some research directions in this field are how to detect (e.g., \cite{moshirpour2012detecting,kjeldaas2021challenges}), how to model (e.g., \cite{hsu2007modeling,mittal2018emergent}), and how to verify and validate (e.g., \cite{brings2018different,brings2020systematic}) the emergent behavior in complex software systems.  

In our previous work \cite{ailane2021toward}, we formalized the definition of emergent behavior in system of systems. The authors discussed various formal and informal definitions and stated some properties of emergent behavior. They used an approach to analyse the development processes in building software systems using an example of a traffic light system. Finally, they gave an informal definition which states that emergent behavior is a non designed behavior which can be beneficial, detrimental or neutral to the system. 

\section{Fundamentals and Formalization} \label{Section:Concepts and definitions}
	
\subsection{Open-/Closed-World Assumption and Circumscription}
The development of software systems and applications can be predicated on different world assumptions: 
In \textit{closed-world assumption} (CWA), the assumption is based on a non-monotonic approach where the idea is that what is not currently known to be true, is false \cite{moore2015context}. A software development under the CWA consists of complete models where anything that is not specified in the model is faulty and wrong. As will be shown in later sections, this is advantageous for developing e.g.,  safety-critical systems where the system is guaranteed (with proofs) to fulfill safety requirements among other requirements.  
On the other hand, the \textit{open-world assumption} (OWA) is based on the alternative reasoning that states: what is predicated in the knowledge base to be true does not affect the not specified information, i.e, what is not know to be true, it is not necessarily false like in CWA. The reasoning of unknown information is independent from what is added to the knowledge base \cite{moore2015context}.

Another non-monotonic reasoning is called the \textit{circumscription}. This type of reasoning can be seen as generalization of the closed-world reasoning where it's creator John McCarthy used it to formalize the common sense assumption that things are as expected unless otherwise specified \cite{mccarthy1980circumscription}.

\subsection{Abstraction Refinement Model}
Abstraction Refinement Model (ARM) is a model of the software evolution process. It consists of three components: descriptions of the software system, a pre-order of relative correctness between system descriptions, and transformations between the system descriptions \cite{keller1993abstraction}. The model is based on the closed-world assumption and a proven model for  building  reliable  and  ``correct  by  construction''  systems. It also provides a balance between forward and reverse engineering in the maintenance phase. In order to define the concept of the ARM, a set of key definitions are provided, which are based on \cite{rausch2001componentware}:

\subsubsection{Syntactic Model} Defines the description techniques (be it textual or graphical) that are available for the system developer to describe and represent the different aspects of the system characteristics (requirements, design, etc.). We use the term model as a short instead of syntactic models.

\subsubsection{Model Semantics}
Reflects the meaning of the syntactic model, and for this reason the semantics are usually preferred to be specified formally to avoid any ambiguity consequences. In this paper, a behavior trace can be viewed as a semantics to a given model.

\subsubsection{Semantic Function}
Is a formal technique to define the relationship between each model and it's corresponding semantics. Hence, the semantic function which we denote as $sem$, is a mapping function that maps each \textit{representation} in the set of models, to an \textit{interpretation} in the set of behaviors. The $sem$ function can be formally defined as follows:
\begin{equation}
sem : M \rightarrow \mathcal{P}(BT)
\end{equation}
\noindent where $M$ is the universe of all models, 
$BT$ is the universe of all behavior traces, 
and $\mathcal{P}(BT)$ is the universe of all behaviors.
$\mathcal{P}$ stands for the power set, the set of all subsets of a set (here: $BT$).

\subsubsection{Evolution} An important aspect during the development process is \textit{evolution}. The evolution  consists of the transformation and change process of the system specifications during the different development phases. At each phase, a new model that defines the system characteristics is developed as can be seen in Fig.~\ref{fig:WFMARM}. Formally, the evolution function \textit{evolve} can be formalized as the following:
\begin{equation}
evolve: M \rightarrow{M}
\end{equation}
Based on the degree of transformation and change of a model $m \in M$ generated from a previous model, an evolution step can be called:

\begin{itemize}
\item \textbf{\textit{Abstraction}}: An evolution is said to be of type \textit{abstraction}, if condition (\ref{AbstractionEquation}) holds. This means that the newly developed model (which we consider as an evolution of the previous developed model) will contain all the semantics of the previous model and may contain more semantics:
\begin{equation}
\label{AbstractionEquation}
 sem(m) \subseteq  sem(evolve(m))
\end{equation}
    \item \textbf{\textit{Refinement}}: An evolution is said to be of type \textit{refinement} if condition (\ref{refinement_constraint}) holds. The semantics of the newly evolved model are all included in the semantics of the previous model. Moreover, the previous developed model may contain more semantics that are not present in the new model. In ARM, this condition must hold between the different models that are generated during the software development process. Once the condition holds, the developer is guaranteed that no extra behavior than specified is to be observed at runtime:
    \begin{equation} \label{refinement_constraint}
    sem(evolve(m)) \subseteq sem(m)
    \end{equation}
\end{itemize}
\begin{itemize}
    \item \textbf{\textit{Total change}}: An evolution is said to be of type \textit{total change} if condition (\ref{TotalChangeEquation}) holds, which indicates that the newly developed model is completely different than the previous model, where no semantics are shared:
    \begin{equation}
    \label{TotalChangeEquation}
    sem(m) \cap  sem(evolve(m)) = \emptyset
\end{equation}

\item \textbf{\textit{Strict evolution}}: An evolution is said to be of type \textit{strict evolution}, if condition (\ref{StrictEvolutionEquation}) holds. This means that the newly evolved model shares some semantics with the previous developed model, however, the previous model may contain a set of semantics that are not part of the newly evolved model and vice versa:
\end{itemize}
\begin{align}
\begin{split}
\label{StrictEvolutionEquation}
 sem(m) \cap sem (evolve(m)) \neq \emptyset \\ 
 \wedge\ sem(m) \nsubseteq  sem (evolve(m)) \\ 
 \wedge\ sem(evolve(m)) \nsubseteq sem(m)\\
 \end{split}
 \end{align}

\section{Extending the abstraction refinement model for systems with emergent behavior}
\label{Section:Extended ARM}

\subsection{ARM application on the Waterfall sequential development process model}

\begin{figure}[!htbp]
\centering
\includegraphics[width=\linewidth]{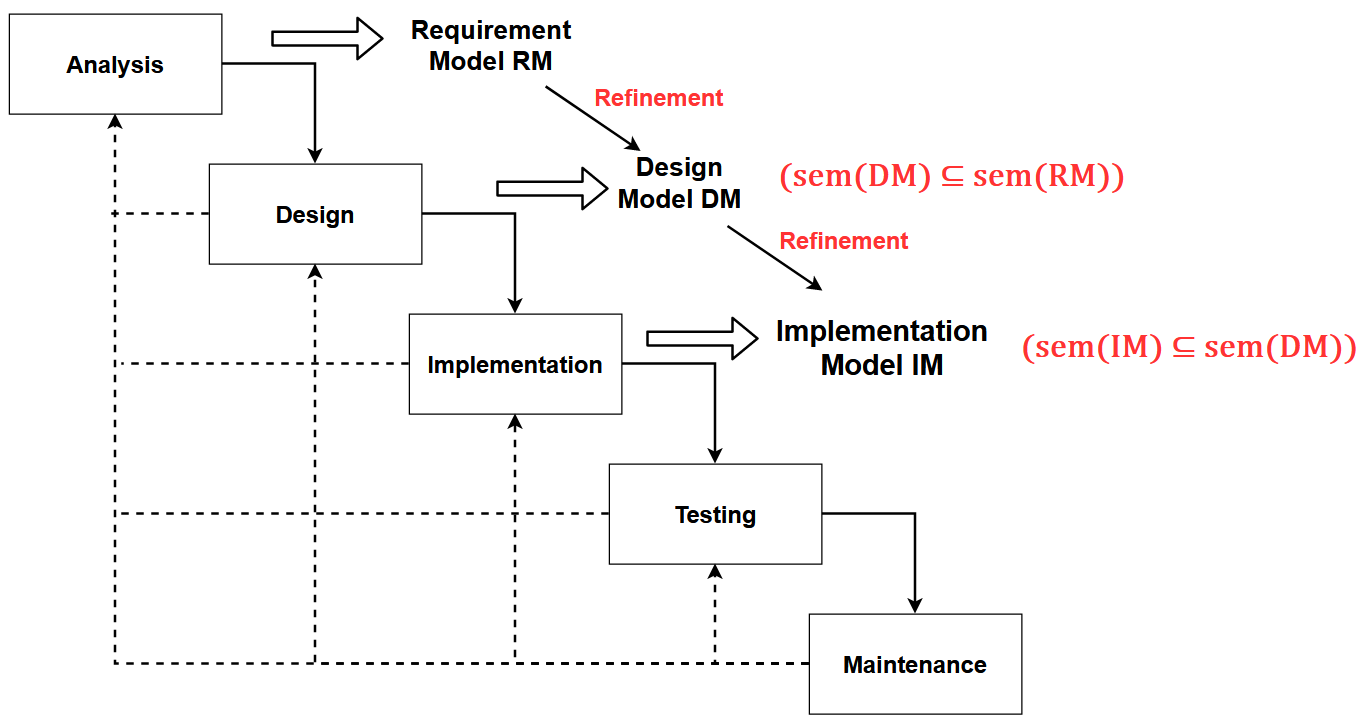}
\caption{Waterfall software development model and refinement according to ARM}
\label{fig:WFMARM}
\end{figure}

Waterfall development process model was introduces early in the seventies  \cite{royce1970software} for methodological development of software systems. The model consists of pre-defined development phases presented in a sequential way in the form of a waterfall (Fig.~\ref{fig:WFMARM}). The software development process usually contains various models described with different notations based on many viewpoints and aspects. 
The completion of a phase results in a new model as an output of the phase, e.g., a requirements model, or a design model.

In order to apply the ARM on the waterfall development process, the following notations are defined and formalized:

Let the set of representations $M$ be the universe of all models and the disjoint subsets corresponding to software development phases as follows:

\begin{itemize}
    \item Universe of all requirement models $RM$ such that:
    $ RM \subseteq M $
    \item Universe of all design models $DM$ such that: 
    $ DM \subseteq M $
    \item Universe of all implementation models $IM$ such that:
    $IM \subseteq M  $
\end{itemize}


Let $BT$ be the universe of all \textit{observable} behavior traces, and the power set of $\mathcal{P}(BT)$ is the universe of all \textit{observable} behaviors.
    
By applying the semantic function $sem$ defined above on a given system $S$, in order to map each model with the respective set of behavior traces, we get the following:
    $sem(rm_s \in RM) := RBT_s \in \mathcal{P}(BT)$;\newline
    $sem(dm_s \in DM) := DBT_s \in \mathcal{P}(BT)$;
    $sem(im_s \in IM) := IBT_s \in \mathcal{P}(BT)$,\newline
\noindent where 
$RBT_s$ is the set of behavior traces of the requirement model $rm$ of a system $S$,
$DBT_s$ is the set of behavior traces of the design model $dm$ of a system $S$, 
$IBT_s$ is the set of behavior traces of the implementation model $im$ of a system $S$.

\begin{figure}[!b]
\centering
\includegraphics[width=0.6\columnwidth]{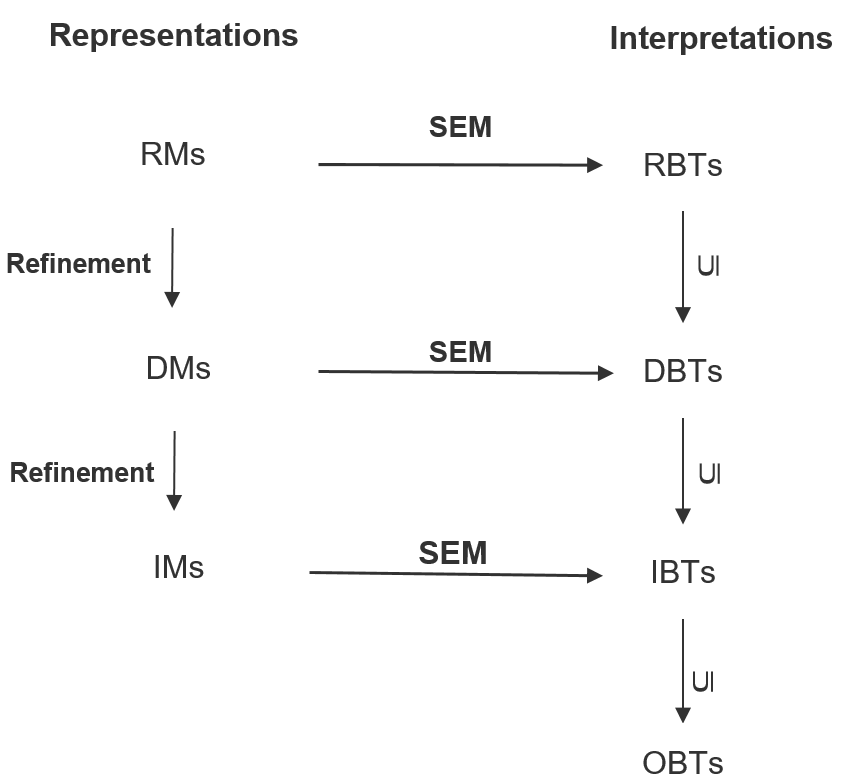}
\caption{Consequences if the refinement constraint of ARM holds during  development}
\label{ARMCWO}
\end{figure}

The basic idea is: if the refinement constraint (equation \ref{refinement_constraint}) holds during the development phases, we are guaranteed to observe a set of behavior traces $OBT_s$ that include no extra behavior than specified (see Fig.~\ref{ARMCWO}). In other words,
ARM guarantees that the following condition always holds:
\begin{equation}\label{observedImplemented}
OBT_s \subseteq IBT_s \subseteq DBT_s \subseteq RBT_s
\end{equation}
%
\subsection{Emergent behavior contradiction in ARM}

In the previous sections, we reviewed the key elements regarding the evolution of different development models and corresponding behaviors of the system during the development process, and how different sets of behavior traces can be mapped to the models to provide an interpretation based on a semantic function. Now, we reason about the nature of the observed behavior and what could be the source and semantics of the emergent behavior, and how does it manifest after deployment at run-time.

First, let $EBT$ be the set of emergent behavior traces, and $EBT_s$ the set of emergent behavior traces of a system $S$ such that:
\begin{equation}
 EBT_s \subseteq EBT
\end{equation}
and
\begin{equation}
 EBT \subseteq \mathcal{P}(BT)
\end{equation}

If we adopt the definition of the emergent behavior presented in \cite{ailane2021toward} where \textit{the emergent behavior is the behavior that has not been specified during development but observed during run-time}, then 
ARM guarantees due to refinement constraints that no emergent behavior will emerge at run-time and the following condition is always guaranteed to take place:
\begin{equation}
    EBT_s \nsubseteq OBT_s \subseteq IBT_s \subseteq DBT_s \subseteq RBT_s 
\end{equation}
This implies that no extra unspecified behavior can be observed at run-time if the refinements constrains hold during the development phases.
On the other hand, and due to real world observation, we have:
\begin{equation}
    EBT_s \subseteq OBT_s \subseteq IBT_s \subseteq DBT_s \subseteq RBT_s 
\end{equation}
Eventually, this leads to a contradiction as the emergent behavior manifest itself to be part of the observed behavior, as it indicates the following:

\begin{equation}\label{EBTsInOBTs}
EBT_s \subseteq OBT_s.    
\end{equation}
and:
\begin{equation} \label{observedNotImplemented}
OBT_s \nsubseteq IBT_s.    
\end{equation}

Both equations \ref{EBTsInOBTs} and \ref{observedNotImplemented} represent the basis observation in \cite{ailane2021toward}, where the following definition of the emergent behavior was provided:

\begin{center}
    \textit{The emergent behavior is the behavior that is not specified during development but appears (observed) at run-time}.
\end{center}

The contradiction of both equations (\ref{observedImplemented}) and (\ref{observedNotImplemented}) is basically due to the closed-world assumption that was assumed by adopting ARM as an evolution model. However, the emergent behavior reflects the fact that the development models are not complete. For this reason we analyse the emergent behavior semantics at each development phase and reason how do such results reflect on the development process:

\noindent \textit{Case 1:} There is an inconsistency between the observed emergent behavior and the required behavior. As shown in equation \ref{StrictEvolutionEquation}, this strict evolution of models says that the emergent behavior can not be fully explained based on the requirement model, this can be formally illustrated as the following:     
\begin{align*}
    sem(rm_s) \cap EBT_s \neq \emptyset  \wedge\ sem(rm_s) \nsubseteq EBT_s  \wedge\ EBT_s \nsubseteq sem(rm_s)
\end{align*}
   \noindent \textit{Case 2:} The emergent behavior is a refinement of the required behavior. However, the emergent behavior indicates that it has been a strict evolution of the design model resulting into an inconsistency between the emergent of the designed behavior, this can be formally illustrated as the following:
   \begin{align*}
        EBTs\ \subseteq sem\ (rm_s)  \wedge\ sem(dm_s) \cap EBT_s \neq \emptyset \\ \wedge\ sem(dm_s) \nsubseteq EBT_s \wedge\ EBT_s \nsubseteq sem(dm_s)
    \end{align*}
   \noindent \textit{Case 3:} The emergent behavior is a refinement of the designed behavior but is not a refinement of the implemented behavior, again the strict evolution of the implementation behavior can only be understood if the implementation behavior consists of something more than the code. If the implementation model is pure code, it is unlikely to imagine that what will be observed would be semantically different than the emergent behavior. However, if the implementation model consists of more than the code (hardware equipment, environment setup,  etc.), it could be the case that the emergent behavior can not be fully explained based on the implementation model, this can be formally illustrated as the following:
   \begin{align*}
        EBT_s\ \subseteq sem\ (dm_s)  \wedge\  sem(im_s) \cap EBT_s \neq \emptyset \\ \wedge\ sem(im_s) \nsubseteq EBT_s \wedge\  EBT_s \nsubseteq sem(im_s)
   \end{align*}

In all cases, a room of an emergent behavior can be found at each development phase, the emergent behavior can be inherited from one phase to another, as well as it can originate at a specific development phase. In the next section, and in order to investigate the emergent behavior, we analyse the specified and unspecified behaviors at each phase to track the notion of the emergent behavior. 
	
\subsection{Extended Abstraction Refinement Model}
\begin{figure}[!htb]
\centering
\includegraphics[width=\linewidth]{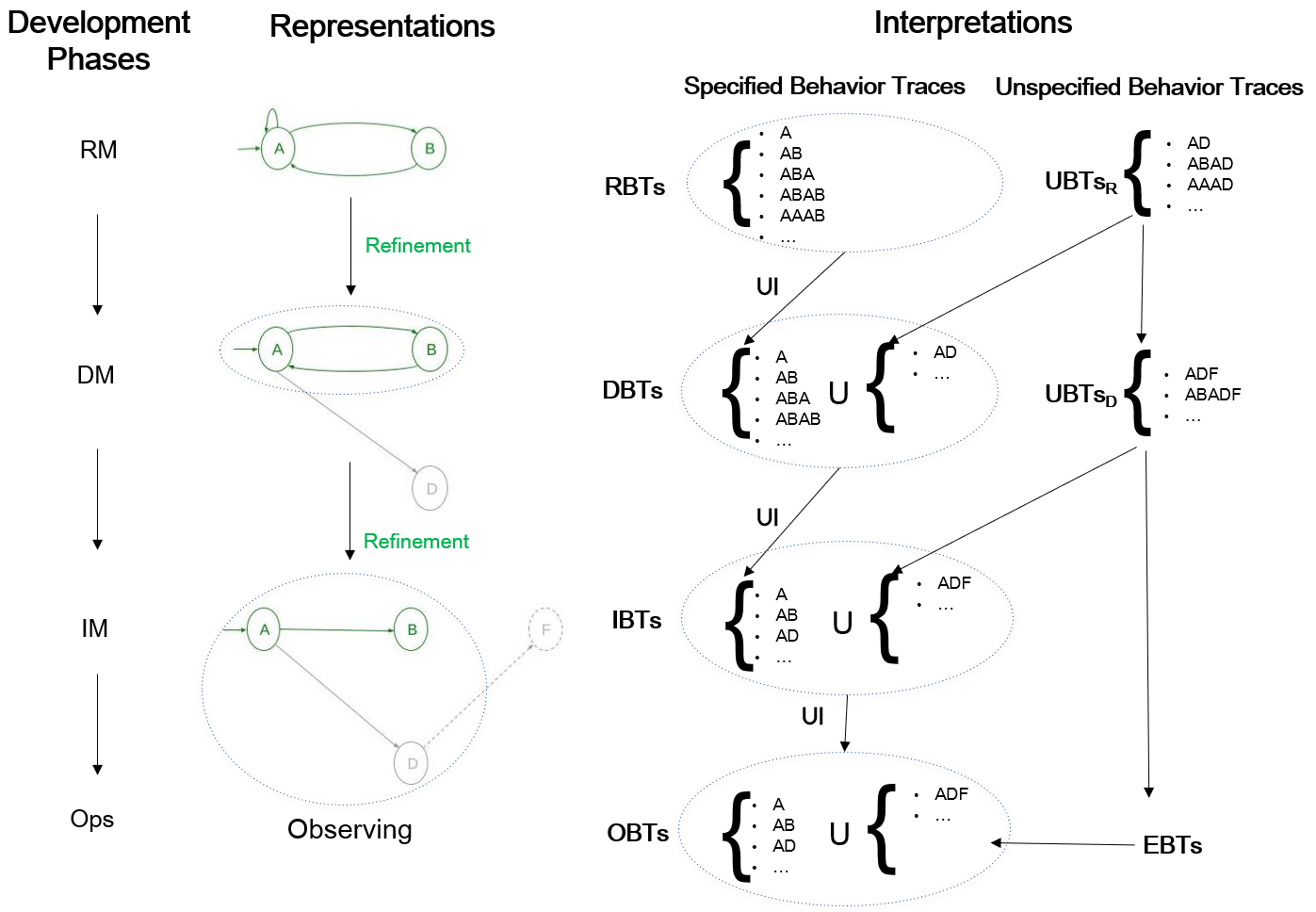}
\caption{Emergent behavior in extended ARM}
\label{Fig:EMARM}
\end{figure}

Based on the definition that the emergent behavior is the behavior that is not specified at development time but is observed at run-time, an investigation of the \textit{unspecified observed behavior traces} of a system $S$ (which we denote as $UOBT_s$) is conducted. At each development phase, we reason about the interpretations in terms of the specified behaviors $RBT_s$, $DBT_s$, $IBT_s$ and the relevant unspecified behaviors: $UOBT_{s_R}$ at a requirement level, and $UOBT_{s_D}$ at a design level, respectively. Here, we assume the implementation model to be pure code, so that no unspecified observed behavior traces exist for the implementation model. 
By extending the development model as shown in Fig.~\ref{Fig:EMARM}, we expect that we will be able to track and identify how the emergent behavior occurs and evolves through the different development phases. 
As explained earlier, the ARM operates based on the notion of the closed world assumption, where the requirement specification is assumed complete. However, in the extended ARM, we reason under an extended-closed world assumption (\textit{circumscription}) which operates on the notion that some intelligent computer systems can reach conclusions based on sparse data, i.e., some requirements can be completed later by analysing data from an initial built system \cite{razniewski2016turning}. Here, we assume that the models are not complete, and the system will be tested to explore extra behaviors and learning specification by observation at run-time.

We interpret that there are specified behavior traces and unspecified behavior traces in each phase. The specified behavior trace is the part that is concrete while the unspecified part is assumed to be behavior traces that can exist in the design and implementation phase but not specified initially in the requirement phase as seen in Fig.~\ref{Fig:EMARM}. In the design and implementation phase, the concrete part which is produced through refinement of the previous phase (each surrounded by the dashed line in Fig.~\ref{Fig:EMARM}) is complete and satisfies the refinement condition while the extra behavior part also relates to the unspecified part of the preceding phase. The union of the concrete part and the extra part is what makes the total specification in each phase. The emergent behavior traces ($EBT_s$) is part of the extra behaviors that can be seen by testing and observing the system during run-time.

\section{Conclusion} \label{sec:conclusion}
Emergent behavior in software systems is the behavior that manifests itself due to incomplete specifications during development. 
To control the emergent behavior throughout the development process we first introduced a formalization based on the ARM as a model of software evolution process.  
We showed that a strict application of the ARM leads to a contradiction, where the emergent behavior can not occur.  
Therefore, we extended the ARM to include the concept of the emergent behavior, where we extend the refinement condition and are now able to deal with incompletenesses in the specifications. 

In a future work, we will apply specification mining techniques to learn specifications at run-time by observation to detect emergent behavior. The approach introduced in this paper will be part of an overall ``DevOps'' development process, where the mined specifications will then be used to automatically complete specifications of the development phases, which should improve in return the understanding and the development of the complex systems. 

\section{Acknowledgement}
The results of this contribution are based on the work of the project “DevOpt: DevOps for Self-Optimizing Emergent Systems”. DevOpt is funded by the Federal Ministry of Education and Research (BMBF) of Germany in the funding programme of “IKT 2020 – Forschung für Innovationen”.

\bibliographystyle{splncs04}
\bibliography{main}

\end{document}